\documentclass{article}

\PassOptionsToPackage{numbers, compress}{natbib}

\usepackage[preprint]{iaseai26}

\usepackage{graphicx}
\usepackage{tabularx,array} 

\usepackage[utf8]{inputenc} 
\usepackage[T1]{fontenc}    
\usepackage{url}            
\usepackage{booktabs}       
\usepackage{amsfonts}       
\usepackage{nicefrac}       
\usepackage{microtype}      
\usepackage{xcolor}         
\usepackage[hidelinks]{hyperref}       

\title{From Awareness to Action: Understanding and Overcoming the Research-Practice Gap in Algorithmic Fairness for Public Health}

\author{%
  Sara Altamirano \\
  Informatics Institute\\
  University of Amsterdam\\
  Science Park 900, 1098 XH Amsterdam\\
  \texttt{s.e.altamirano@uva.nl}
  \And
  Tijs Portegies \\
  Informatics Institute\\
  University of Amsterdam\\
  Science Park 900, 1098 XH Amsterdam\\
  \texttt{t.c.portegies@uva.nl}
  \And
  Sennay Ghebreab \\
  Informatics Institute\\
  University of Amsterdam\\
  Science Park 900, 1098 XH Amsterdam\\
  \texttt{s.ghebreab@uva.nl}
}

\begin{document}

\maketitle

\begin{abstract}
Algorithmic fairness is essential for responsible ML-driven public health research, yet its practical implementation remains limited. To investigate this awareness–action gap, we conducted a sequential mixed-methods study comprising expert interviews, an online survey, and systematic mapping. The expert interviews informed the design of the survey, which in turn revealed fragmented definitions of fairness, limited training and guidance, reliance on external sources, and rare use of formal assessment, mitigation, or monitoring. These findings were subsequently mapped onto three established research–practice gap lenses: the Knowledge–Practice Gap, the Knowledge-to-Action Cycle, and the Knowing–Doing Gap, each offering complementary perspectives. Building on this synthesis, we introduce the Fairness-to-Action framework, which integrates methodological, organizational, and systemic dimensions to identify where translation of algorithmic fairness knowledge stalls. Our analysis shows that fairness remains weakly institutionalized, translation mechanisms are externally driven, and system-level priorities continue to emphasize accuracy over fairness. These insights suggest critical leverage points for advancing safe, fair, and ethical ML-driven public health research practice.
\end{abstract}

\section{Introduction}

Algorithmic decision-making (ADM) is reshaping public health through applications in disease surveillance, outcome prediction, resource allocation, and identification of vulnerable groups \citep{benke_artificial_2018, morgenstern_predicting_2020, wiemken_machine_2020, mhasawade_machine_2021, olawade_using_2023, jungwirth_artificial_2023}. By integrating large-scale data and predictive modeling, machine learning (ML) enables new forms of population-level monitoring and intervention. Yet these opportunities carry significant risks: models trained on incomplete or skewed data can reproduce socioeconomic, racial, or geographic disparities \citep{obermeyer_dissecting_2019, fletcher_addressing_2021, flores_addressing_2024}. In public health, such risks can be amplified because determinants of health are multi-layered, trade-offs across groups are often unavoidable, and decisions often span multiple institutions \citep{wesson_risks_2022, tsai_algorithmic_2022, chin_guiding_2023}. Technical validity alone is therefore insufficient to ensure safe or equitable outcomes \citep{thomasian_advancing_2021, nazer_bias_2023}, making fairness a central pillar of responsible ML.

Accordingly, algorithmic fairness has emerged as a central ethical and societal concern. Approaches range from quantitative criteria such as demographic parity \citep{feldman_certifying_2015} and equalized odds \citep{hardt_equality_2016} to procedural measures emphasizing transparency and stakeholder participation \citep{dwork_fairness_2012, barocas_fairness_2023, caton_fairness_2024}. Growing awareness reflects a shift from predictive performance toward equity and accountability as conditions for responsible ML. Yet awareness alone rarely translates into systematic mitigation or transparent reporting, and practices remain fragmented across domains \citep{holstein_improving_2019, madaio_co-designing_2020, chen_unmasking_2024}. Even in advanced health systems, fairness research often remains disconnected from public health implementation \citep{thomasian_advancing_2021, wesson_risks_2022}. This persistent gap limits the real-world impact of fairness and calls for empirical insight into how researchers themselves interpret and apply it.

To address this gap, we examine how researchers in ML-driven public health \textit{perceive} and \textit{operationalize} fairness across the ML lifecycle: conceptualization, awareness, evaluation, design, and application (adapted from \citep{royce_managing_1970, molenda_search_2015, suresh_framework_2021}). These five dimensions structure our research sub-questions (RSQ1 to RSQ5). To interpret the observed disconnect between fairness research and practice, we draw on established theories from implementation science: the Knowledge–Practice Gap \citep{grimshaw_knowledge_2012}, the Knowledge-to-Action (KTA) Cycle \citep{graham_lost_2006}, and the Knowing–Doing Gap \citep{pfeffer_knowing-doing_2000}. Using a sequential mixed-methods design, we first conducted expert interviews to inform the design of an online survey, then analyzed survey responses to map researcher perceptions and practices against these frameworks. This synthesis yields the Fairness-to-Action (F2A) framework, which identifies where fairness knowledge stalls before reaching practice and outlines conditions for more effective translation in ML-driven public health research. We ground the study in the Netherlands, a salient case as it combines mature public health and ML capacity, advanced data infrastructures, and an active artificial intelligence (AI) governance landscape with documented inequities among ethnic minorities and socioeconomically disadvantaged groups \citep{ikram_disease_2014, kroneman_netherlands_2016, ilozumba_ethnic_2022}. These features make it an informative case with lessons for high-resource settings.

\paragraph{Statement of Contributions.}
First, we empirically examine how researchers in ML-driven public health perceive and operationalize algorithmic fairness across the ML lifecycle through an interview-informed survey. Second, we integrate evidence with implementation science by situating the observed awareness–action gap within three established theoretical lenses, identifying pathways for improving fairness adoption (see Section~\ref{sec:methods}). Third, we extend this theoretical synthesis to the fairness domain and introduce the F2A framework (see Section~\ref{sec:results}), which can help explain how fairness knowledge becomes sustained practice and where the process often stalls. Together, these contributions synthesize empirical and theoretical insights to advance safe, fair, and ethical ML-driven public health research practice.

\section{Related Work}

Reviews of ML-driven public health consistently identify sources of bias and equity risks, ranging from data absenteeism and representational gaps to mis-specified objectives and model misuse \citep{mhasawade_machine_2021, wesson_risks_2022, berdahl_strategies_2023, chen_unmasking_2024, flores_addressing_2024}. Empirical studies further demonstrate subgroup disparities and performance variation: racial bias in commercial risk scores \citep{obermeyer_dissecting_2019}, underdiagnosis in underserved populations \citep{seyyed-kalantari_underdiagnosis_2021}, poor external generalizability \citep{zech_variable_2018}, gender imbalance in imaging datasets \citep{larrazabal_gender_2020}, and inequities embedded in race corrections \citep{vyas_hidden_2020}. In response, mitigation strategies emphasize fairness metrics, explainability, and lifecycle-based awareness \citep{chen_ethical_2021, grote_algorithmic_2022, suresh_framework_2021, nazer_bias_2023, thomasian_advancing_2021}, complemented by structured reporting and evaluation standards \citep{cruz_rivera_guidelines_2020, liu_reporting_2020, hernandez-boussard_minimar_2020, collins_tripodai_2024, moons_probastai_2025, wawira_gichoya_equity_2021}. Conceptual work stresses transparency, accountability, and inclusion as pillars for algorithmic fairness \citep{sikstrom_conceptualising_2022}, yet metrics and documentation alone are insufficient: over-reliance may reinforce inequities \citep{mccradden_ethical_2020, chin_guiding_2023, fletcher_addressing_2021}, particularly where population-level trade-offs, structural determinants, and fragmented governance complicate evaluation and mitigation \citep{panteli_artificial_2025}. Collectively, this literature underscores the limits of technical fixes and highlights the need for integrated governance, participatory design, and context-sensitive evaluation. 
Despite extensive technical and policy work, little is known about how researchers themselves engage with algorithmic fairness. Increasingly, fairness in ML-driven public health is framed as a socio-technical concern. Ethical and policy analyses emphasize justice, transparency, and inclusion \citep{aysolmaz_public_2023, sikstrom_conceptualising_2022, berdahl_strategies_2023, wesson_risks_2022}, while human-computer interaction and social science studies introduce tools such as model cards, datasheets, and checklists \citep{mitchell_model_2019, gebru_datasheets_2021, madaio_co-designing_2020}. However, adoption remains inconsistent. Empirical studies reveal heterogeneity: industry practitioners report limited uptake of fairness practices \citep{holstein_improving_2019}, AI developers cite the lack of fair data, guidelines, and expertise as barriers \citep{vorisek_artificial_2023}, and systematic reviews show fragmented definitions and context-dependent judgments \citep{starke_fairness_2022, kasun_academic_2024}. Broader critiques link fairness to structural inequities and governance gaps \citep{mccradden_ethical_2020, chin_guiding_2023, kim_heaal_2024}. In the Netherlands, these global trends are reflected in a context where advanced data infrastructures coexist with persistent inequities among ethnic minorities and disadvantaged groups \citep{ikram_disease_2014, kroneman_netherlands_2016, ilozumba_ethnic_2022}. Despite policy initiatives such as the Algorithm Register \citep{noauthor_algorithm_nodate} and revised data classification standards \citep{CBS_reform_2022}, explicit fairness evaluations in ML-driven public health remain rare \citep{altamirano_machine_2025}. This study extends the literature by centering researchers as the unit of analysis and examining the research–practice gap through an interview-informed survey.

\section{Theoretical Framework}
\label{sec:theoretical-framework}

Translating algorithmic fairness into practice requires attention to methodological, organizational, and systemic factors. Implementation science offers complementary perspectives for understanding how knowledge moves, or fails to move, into practice \citep{fixsen2005_implementation, greenhalgh_diffusion_2004, proctor_outcomes_2011, peters_trends_2022}; F2A integrates these lenses to interpret where fairness translation falters. This approach is particularly relevant to public health, where translation spans multiple institutions, competing priorities, and unavoidable trade-offs across populations, underscoring the need to assess both implementation processes and outcomes.

To structure our methodological approach, we draw on three established theories that together specify the \textit{what}, \textit{how}, and \textit{why} of fairness translation. The Knowledge–Practice Gap \citep{grimshaw_knowledge_2012} delineates what characterizes the current state of the gap: effective knowledge rarely becomes routine practice without targeted translation activities and supportive conditions. The KTA Cycle \citep{graham_lost_2006} clarifies how fairness knowledge is transferred through an action cycle that adapts evidence to context, addresses barriers, implements strategies, and evaluates sustainment. The Knowing–Doing Gap \citep{pfeffer_knowing-doing_2000} explains why the gap persists, emphasizing organizational dynamics that substitute discourse for action, privilege narrow metrics, or sustain risk-averse norms. These theories have been extended through models such as the Consolidated Framework for Implementation Research (CFIR), the Active Implementation Frameworks, and the Centers for Disease Control and Prevention (CDC) KTA guide \citep{damschroder_fostering_2009, fixsen2005_implementation, centers_for_disease_control_and_prevention_applying_2014}. Adaptations emphasize team-based learning, continuous feedback, and iterative improvement \citep{kilo_framework_1998, ivers_2025}.

In short, these perspectives clarify \textit{what} defines the gap, \textit{how} fairness knowledge is transferred or stalled, and \textit{why} it persists. This threefold structure provides a systems-level lens: methodological factors shape how knowledge is produced, organizational factors govern its use, and systemic factors determine its sustainment. Together, these dimensions offer a foundation for analyzing fairness translation across the ML lifecycle. Section~\ref{sec:methods} outlines how we applied these lenses.

\section{Methods}
\label{sec:methods}

\subsection{Empirical Study}

\begin{figure}[t]
  \centering
  \includegraphics[width=\linewidth]{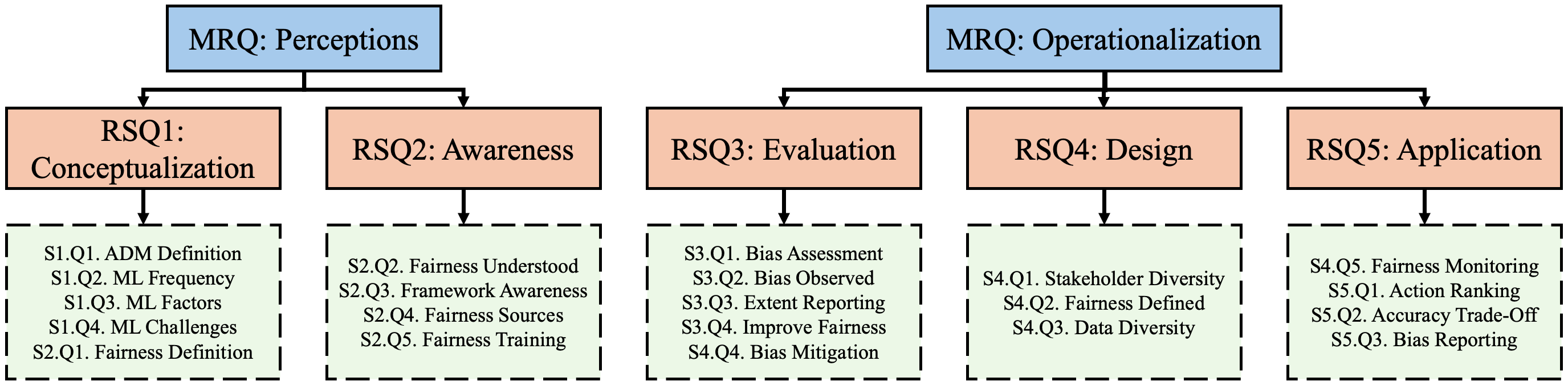}
  \caption{Mapping of survey items to Main Research Question (MRQ) and Research Sub-Questions (RSQs). Solid borders represent research questions, dashed borders indicate grouped items.}
  \label{fig:rsq-mapping}
\end{figure}

We used a two-phase sequential exploratory mixed-methods design \citep{creswell_qualitative_2018} to examine how algorithmic fairness is perceived and operationalized in ML-driven public health. Phase one comprised semi-structured expert interviews (28 questions across four sections on ADM use, fairness awareness and practice, and study-level practices adapted from \citep{gebru_datasheets_2021, madaio_co-designing_2020, lekadir_future-ai_2024}). A thematic analysis grounded constructs, aligned practitioner vocabulary, and shaped specific prompts for the survey. Instruments were piloted (interview guide with two colleagues; survey with five) with edits limited to wording and logic. Survey items were derived from coded interview themes to ensure content validity and traceability to qualitative constructs, supported by expert triangulation during iterative revision. Item to RSQ mapping is shown in Figure~\ref{fig:rsq-mapping}. Reporting follows CROSS \citep{sharma_consensus-based_2021} and CHERRIES \citep{eysenbach_improving_2004}. Full survey questionnaire can be found in Appendix \ref{appendix_full_survey}; additional documentation is available on request.

Interviewees were recruited purposively and via snowballing. Of 33 invitations, four experts participated, meeting two criteria: willingness to reflect on fairness and demonstrated expertise. Two interviews were conducted via recorded video call and two via written questionnaires; all were in English and lasted 30 to 45 minutes. Participants represented diverse academic, policy, and technical backgrounds, spanning varying levels of seniority and institutional roles. While thematic saturation was not sought, the interviews were used for instrument development and conceptual coverage in an exploratory design. The survey targeted researchers from Dutch institutions working at the ML–public health interface, identified through authorship records, professional networks, and public listings; interviewees were excluded. We distributed 105 personalized invitations and supplemented outreach via social media, academic events, and word of mouth, so the total reach exceeded 105. The study used a cross-sectional online survey administered in Qualtrics (English, February–August 2025) via a single anonymous link. We obtained 73 valid responses, defined as submissions with verifiable institutional email addresses and $\ge$90\% completion. Given the lack of a complete sampling frame for this niche researcher population, we treat the data as exploratory and use them to identify recurring patterns rather than estimate prevalence. A filter question excluded respondents without ML experience, and the estimated completion time was 15 minutes. Data-quality controls included required consent, completeness checks, and a post hoc email check to limit duplicates; participants could revise responses before submission. Respondents could enter a raffle for five \texteuro10 gift cards; contact details were used solely for validation and prize notification, stored separately from survey data, and deleted after use.

Quantitative analysis included descriptive statistics, frequency distributions, and cross-tabulations by S0.Q1 (Primary Field), S0.Q3 (Years Experience), and S0.Q5 (Institution Type). Multiple-answer items were summarized as counts and percentages; the ranking item S5.Q1 by mean rank and rank-1 frequency; Likert items by category counts and percentages, with occasional grouping for interpretability as reported in Section~\ref{sec:results}. The survey included multiple choice, Likert-scale, ranking, and open-text items. Inferential testing was not performed because subgroup counts fell below thresholds for valid $\chi^2$ or regression analyses; descriptive reporting was emphasized instead. Data cleaning removed direct identifiers and high-risk pseudo-identifiers, deleted linkage keys, validated institutional email domains, and removed emails flagged by the post hoc check. The resulting dataset is anonymized with no retained keys. Item non-response was lower than 1\%; no imputation was performed. All analyses were performed in Python (v3.13).

Open-text responses were analyzed using a keyword-assisted inductive approach. A structured codebook defined six themes with descriptions and keyword lists. Two coders independently assigned each response to a single theme based on keyword matches and human judgment; when no keywords matched, the main conceptual idea guided coding. Disagreements were resolved through discussion until consensus. Coders were blinded to respondent metadata. Intercoder reliability statistics were not calculated; iterative consensus coding was used instead. Category frequencies are reported; codebooks with themes, descriptions, and keywords are provided in Appendix~\ref{tab:codebook_adm}.

Separate consent was obtained for interviews and surveys. The study involved no sensitive data or vulnerable populations, followed University of Amsterdam ethical guidance and GDPR, and allowed withdrawal at any time. See Appendix~\ref{appendix:ethics} for details.

\subsection{Theoretical Embedding}

To examine fairness translation mechanisms, survey items were embedded within the three frameworks presented in Section~\ref{sec:theoretical-framework}, with F2A used as an analytical scaffold rather than proposing a new theoretical model. Items were mapped along three axes: \textit{what} fairness knowledge exists or is missing \citep{grimshaw_knowledge_2012}, \textit{how} fairness knowledge is transferred and where translation stalls \citep{graham_lost_2006}, and \textit{why} organizational mechanisms sustain the research-practice gap \citep{pfeffer_knowing-doing_2000}. Framework constructs were adapted into fairness-specific diagnostic questions to guide mapping. Two researchers independently coded items using a structured spreadsheet; discrepancies were resolved through discussion, and coding decisions were retained for traceability. During interpretation, convergence across frameworks and framework-specific divergences were identified, and subgroup patterns by field, institution type, and ML experience were assessed.

The outputs of this process were organized into three framework-specific tables that present survey findings in terms of the adapted diagnostic questions. Integration was conducted at the interpretive stage by comparing convergence and divergence across the three frameworks. This approach allowed us to preserve the integrity of each framework while highlighting their complementary insights into the fairness translation gap. Full item-to-framework mappings, including coder notes, are available upon request.

\section{Results}
\label{sec:results}
\subsection{Empirical Study}

\paragraph{Sample size considerations.}
Our final sample consisted of 73 respondents. Results are presented using descriptive statistics,
cross-tabulations where relevant, and illustrative excerpts from open-ended responses. Percentages are descriptive indicators of patterns in this sample, not population estimates. Detailed distributions for all survey items from RSQ1 to RSQ5 appear in
Tables~\ref{tab:rsq1_conceptualization} to \ref{tab:rsq5_application}.

\paragraph{Respondent Characteristics: Who Are the Researchers?}
We analyzed 73 valid responses ($\geq$90\% completion, verified contacts). As summarized in Table~\ref{tab:respondent_characteristics}, respondents were from technical and applied fields, with academics the largest role group and universities the most common affiliation. Early-career researchers were concentrated in universities, while those with $\geq$4 years of experience were more often in hospitals and government institutes. Overall, the majority of respondents reported more than four years of experience, with smaller numbers affiliated to the private sector and non-governmental organizations (NGOs).

\begin{table}[h]
\centering
\caption{Characteristics of survey respondents.}
\label{tab:respondent_characteristics}
\begingroup
\scriptsize 
\setlength{\tabcolsep}{2pt}
\renewcommand{\arraystretch}{1.05}

\begin{tabular}{@{}p{2.6cm}p{1cm}p{2.2cm}p{1cm}p{1.8cm}p{1cm}p{2.2cm}p{1cm}@{}}
\toprule
\textbf{Primary field} & $n$ (\%) & \textbf{Current role} & $n$ (\%) & \textbf{Years experience} & $n$ (\%) & \textbf{Institution type} & $n$ (\%) \\
\midrule
DS and ML & 23 (32) & Academic/Researcher & 39 (53) & $<$1 year & 2 (3) & University & 30 (41) \\
PH Research & 21 (29) & PH Professional & 14 (19) & 1--3 years & 21 (29) & Hospital & 20 (27) \\
Epidemiology/Biostatistics & 15 (21) & DS/ML Engineer & 12 (16) & 4--7 years & 27 (37) & Government & 16 (22) \\
PH Policy/Management & 8 (11) & Policymaker & 7 (10) & 8--10 years & 11 (15) & NGO & 6 (8) \\
Other\textsuperscript{1} & 6 (8) & Other\textsuperscript{2} & 1 (1) & $>$10 years & 12 (16) & Private sector & 1 (1) \\
\bottomrule
\end{tabular}

\begin{flushleft}
\scriptsize
\textsuperscript{1} Includes: Medicine; Neuroscience; Sexual and Reproductive Health and Rights/digital rights; Atmospheric Sciences; (Biological) Psychology; Mathematics/Computational Science. \\
\textsuperscript{2} Civil society organization (advisor). \\
Abbreviations: DS, data science; ML, machine learning; PH, public health; Govt., government; NGO, non-governmental organization. \\
Percentages are rounded and may not total exactly 100\%.
\end{flushleft}
\endgroup
\end{table}

\subsubsection*{RSQ1. Conceptualization: How do ML-driven public health researchers understand and engage with algorithmic decision-making (ADM) and fairness?}

Respondents described ADM mainly as decision support or data-based modeling, emphasizing algorithms that assist expert judgment or derive predictions from data, for example, ``\textit{Building AI-driven tools to assist decision-making.}'' Algorithmic fairness was most often defined as equal treatment or equal performance across groups, typically expressed as ``\textit{Ensuring models do not underperform for minority groups.}'' In (S1.Q2 to S1.Q4), ML use was frequent among data scientists but less common in public health; data quality and availability were the most cited enablers, while data interoperability and limited ML expertise were leading barriers. In general, ADM and fairness were interpreted through disciplinary lenses: technical for data scientists; ethical and practical for public health researchers, reflecting uneven familiarity and institutional constraints related to roles, data access, and governance.

\begin{table}[h]
\centering
\caption{RSQ1: Conceptualization, response distribution from survey items S1.Q1–S1.Q4 and S2.Q1.}

\label{tab:rsq1_conceptualization}
\begingroup
\scriptsize
\setlength{\tabcolsep}{0pt}

\begin{tabular}{@{}p{1.6cm}p{1cm}p{2.1cm}p{1cm}p{1.3cm}p{1cm}p{1.8cm}p{1cm}p{1.95cm}p{1cm}@{}}
\toprule
\multicolumn{2}{c}{\parbox[c]{2.6cm}{\centering\textbf{(S1.Q1) ADM Definition}}} &
\multicolumn{2}{c}{\parbox[c]{3.1cm}{\centering\textbf{(S2.Q1) Fairness Definition}}} &
\multicolumn{2}{c}{\parbox[c]{2.3cm}{\centering\textbf{(S1.Q2) ML Frequency}}} &
\multicolumn{2}{c}{\parbox[c]{2.8cm}{\centering\textbf{(S1.Q3) ML Factors}}} &
\multicolumn{2}{c}{\parbox[c]{2.95cm}{\centering\textbf{(S1.Q4) ML Challenges}}} \\
\cmidrule(lr){1-2}\cmidrule(lr){3-4}\cmidrule(lr){5-6}\cmidrule(lr){7-8}\cmidrule(l){9-10}
\textit{Answers} & \textit{n (\%)} &
\textit{Answers} & \textit{n (\%)} &
\textit{Answers} & \textit{n (\%)} &
\textit{Answers} & \textit{n (\%)} &
\textit{Answers} & \textit{n (\%)} \\
\midrule
Decision Supp. & 22 (30) &
Equal Treatment & 21 (29) &
Rarely & 1 (1) &
Data Q\&A & 62 (85) &
Data access & 50 (68) \\
Data Analysis & 14 (19) &
Protected Attributes & 20 (27) &
Occasionally & 14 (19) &
Model costs & 35 (48) &
Computing power & 29 (40) \\
Model Focus & 13 (18) &
Outcome Fairness & 14 (19) &
Sometimes & 13 (18) &
Interoperability & 31 (42) &
Legal/Regulatory & 30 (41) \\
Pattern Recog. & 10 (14) &
Standards/Processes & 10 (14) &
Frequently & 21 (29) &
Transparency & 24 (33) &
Interoperability & 52 (71) \\
Rules Based & 5 (7) &
Context \& Account. & 0 (0) &
Extensively & 24 (33) &
Ethics & 13 (18) &
ML expertise & 39 (53) \\
Governance & 4 (5) &
Equity \& Disparities & 5 (7) &
-- & -- &
Regulatory reqs & 24 (33) &
Transparency & 20 (27) \\
Unclassified & 5 (7) &
Unclassified & 3 (4) &
-- & -- &
Other\textsuperscript{1} & 8 (11) &
Other\textsuperscript{2} & 2 (3) \\
\midrule
\multicolumn{2}{c}{{\textit{Total: }73 (100\%)}} &
\multicolumn{2}{c}{{\textit{Total: }73 (100\%)}} &
\multicolumn{2}{c}{{\textit{Total: }73 (100\%)}} &
\multicolumn{2}{c}{{\textit{Base: }N=73 (mult. answer)}} &
\multicolumn{2}{c}{{\textit{Base: }N=73 (mult. answer)}} \\
\bottomrule
\end{tabular}
\begin{flushleft}
\scriptsize
\textsuperscript{1} “Other” (S1.Q3): trade-off explainability vs accuracy, implementation limits, complex or high-dimensional data, limited support. \textsuperscript{2}~“Other” (S1.Q4): evaluating personalized causal effects or combined challenges. Percentages are rounded and may not total exactly 100\%. Abbreviations: Supp., Support; Recog., Recognition; Account., Accountability; Q\&A, quality \& availability; reqs., requirements; mult., multiple.
\end{flushleft}
\endgroup
\end{table}

\subsubsection*{RSQ2. Awareness: How familiar are researchers with algorithmic fairness in ML-driven public health research?}

Awareness of algorithmic fairness was generally low. In (S2.Q2) Fairness Understood, 79\% reported little to moderate familiarity, and only one in five indicated strong or full understanding. In (S2.Q3) Framework Awareness, 73\% were aware of existing frameworks but without concrete details, while the remainder were unsure or believed none exist; awareness within organizations appeared mainly in hospitals. In (S2.Q4) Fairness Sources, most relied on academic or professional channels (e.g., peer-reviewed articles, networks, and conferences) whereas 37\% did not stay updated and none cited internal training. In (S2.Q5) Fairness Training, about 70\% reported no or minimal learning opportunities, with structured programs rare outside hospitals. In general, fairness knowledge was uneven and concentrated in academic settings, reflecting unequal institutional capacity and limited access to formal training.

\begin{table}[h!]
\centering
\caption{RSQ2: Awareness, response distribution from survey items S2.Q2–S2.Q5.}
\label{tab:rsq2_awareness}
\begingroup
\scriptsize
\setlength{\tabcolsep}{0pt} 

\begin{tabular}{@{}p{2cm}p{1.2cm}p{2.3cm}p{1.2cm}p{2.5cm}p{1.2cm}p{2.1cm}p{1.20cm}@{}}
\toprule
\multicolumn{2}{c}{\parbox[c]{3.2cm}{\centering\textbf{(S2.Q2) Fairness Understood}}} &
\multicolumn{2}{c}{\parbox[c]{3.25cm}{\centering\textbf{(S2.Q3) Framework Awareness}}} &
\multicolumn{2}{c}{\parbox[c]{3.25cm}{\centering\textbf{(S2.Q4) Fairness Sources}}} &
\multicolumn{2}{c}{\parbox[c]{3.25cm}{\centering\textbf{(S2.Q5) Fairness Training}}} \\
\cmidrule(lr){1-2}\cmidrule(lr){3-4}\cmidrule(lr){5-6}\cmidrule(l){7-8}
\textit{Answers} & \textit{n (\%)} &
\textit{Answers} & \textit{n (\%)} &
\textit{Answers} & \textit{n (\%)} &
\textit{Answers} & \textit{n (\%)} \\
\midrule
Not at all & 2 (3) &
Yes, within group/org & 7 (10) &
Academic networks & 26 (36) &
Regular training & 2 (3) \\
Minimally & 31 (43) &
Only in field & 17 (23) &
Peer-reviewed articles & 34 (47) &
Periodic workshops & 9 (12) \\
Moderately & 25 (34) &
Aware: no details & 29 (40) &
Conferences/workshops & 18 (25) &
External training & 13 (18) \\
Strongly & 11 (15) &
None exist & 6 (8) &
Social media/online & 20 (27) &
Plan to start & 6 (8) \\
Fully & 4 (5) &
Not sure & 11 (15) &
Internal training & 0 (0) &
No programs & 37 (51) \\
-- & -- &
Other\textsuperscript{1} & 3 (4) &
Do not stay updated & 27 (37) &
Other\textsuperscript{3} & 6 (8) \\
-- & -- & -- & -- &
Other\textsuperscript{2} & 2 (3) &
-- & -- \\
\midrule
\multicolumn{2}{c}{\textit{Total: }73 (100\%)} &
\multicolumn{2}{c}{\textit{Total: }73 (100\%)} &
\multicolumn{2}{c}{\textit{Base: }N=73 (multiple answer)} &
\multicolumn{2}{c}{\textit{Total: }73 (100\%)} \\
\bottomrule
\end{tabular}

\begin{flushleft}
\scriptsize
\textsuperscript{1} “Other” (S2.Q3): limited or no awareness of organizational fairness guidelines; one cited public resource. \textsuperscript{2} “Other” (S2.Q4): fairness discussed informally within teams or during peer review. \textsuperscript{3} “Other” (S2.Q5): uncertainty or lack of knowledge about existing training or standards. Percentages are rounded and may not total exactly 100\%. Abbreviations: org., organization.
\end{flushleft}

\endgroup
\end{table}

\subsubsection*{RSQ3. Evaluation: How do researchers assess and report algorithmic bias in ML-driven public health research?}

Evaluation of algorithmic bias remained largely informal and descriptive. In (S3.Q1) Bias Assessment, most relied on subgroup sensitivity or demographic analyses and informal discussion, while formal approaches such as fairness metrics or external review were rare. In (S3.Q2) Bias Observed, reported issues mainly concerned prediction accuracy gaps and data quality or sample diversity, with one fifth reporting none. In (S3.Q3) Extent Reporting, 75\% described minimal or moderate reporting of bias. In (S3.Q4) Improve Fairness, actions centered on open science and transparency, such as code sharing, algorithm availability, and stakeholder involvement, while formal accountability mechanisms were uncommon. In (S4.Q4) Bias Mitigation, 72\% reported none or minimal mitigation. In general, bias evaluation practices were ad hoc and descriptive rather than methodological.

\begin{table}[h]
\centering
\caption{RSQ3: Evaluation, response distribution from survey items S3.Q1–S3.Q4 and S4.Q4.}
\label{tab:rsq3_evaluation}
\begingroup
\scriptsize
\setlength{\tabcolsep}{0pt}

\begin{tabular}{@{}p{2.1cm}p{1cm}p{2.15cm}p{1cm}p{1.1cm}p{1cm}p{2.3cm}p{1cm}p{1.1cm}p{1cm}@{}}
\toprule
\multicolumn{2}{c}{\parbox[c]{3.1cm}{\centering\textbf{(S3.Q1) Bias Assessment}}} &
\multicolumn{2}{c}{\parbox[c]{3.15cm}{\centering\textbf{(S3.Q2) Bias Observed}}} &
\multicolumn{2}{c}{\parbox[c]{2.1cm}{\centering\textbf{(S3.Q3) Extent Reporting}}} &
\multicolumn{2}{c}{\parbox[c]{3.3cm}{\centering\textbf{(S3.Q4) Improve Fairness}}} &
\multicolumn{2}{c}{\parbox[c]{2.1cm}{\centering\textbf{(S4.Q4) Bias Mitigation}}} \\
\cmidrule(lr){1-2}\cmidrule(lr){3-4}\cmidrule(lr){5-6}\cmidrule(lr){7-8}\cmidrule(l){9-10}
\textit{Answers} & \textit{n (\%)} &
\textit{Answers} & \textit{n (\%)} &
\textit{Answers} & \textit{n (\%)} &
\textit{Answers} & \textit{n (\%)} &
\textit{Answers} & \textit{n (\%)} \\
\midrule
Sensitivity analysis & 49 (67) &
Accuracy gaps & 32 (44) &
None & 3 (4) &
Code sharing & 45 (62) &
Not at all & 33 (45) \\
Demographic vars. & 39 (53) &
Data quality issues & 29 (40) &
Minimal & 38 (52) &
Algorithm availability & 41 (56) &
Minimal & 20 (27) \\
Informal discussion & 38 (52) &
Sample diversity lim. & 23 (32) &
Moderate & 17 (23) &
Stakeholders involved & 32 (44) &
Moderate & 6 (8) \\
Fairness metrics & 10 (14) &
Underrepresentation & 21 (29) &
Strong & 9 (12) &
Audit reports & 7 (10) &
Strong & 13 (18) \\
External review & 8 (11) &
No bias observed & 16 (22) &
Full & 1 (1) &
Other\textsuperscript{2} & 5 (7) &
Full & 1 (1) \\
No measures & 9 (12) &
Other\textsuperscript{1} & 3 (4) &
-- & -- &
-- & -- &
-- & -- \\
\midrule
\multicolumn{2}{c}{\textit{Base: }N=73 (multiple answer)} &
\multicolumn{2}{c}{\textit{Base: }N=73 (multiple answer)} &
\multicolumn{2}{c}{\textit{Total: }73 (100\%)} &
\multicolumn{2}{c}{\textit{Base: }N=73 (multiple answer)} &
\multicolumn{2}{c}{\textit{Total: }73 (100\%)} \\
\bottomrule
\end{tabular}
\begin{flushleft}
\scriptsize
\textsuperscript{1} “Other” (S3.Q2): not directly assessing fairness or limited relevance to population analyses. \textsuperscript{2} “Other” (S3.Q4): references to internal meetings, peer-reviewed publications, or new research projects. Percentages are rounded and may not total exactly 100\%. Abbreviations: vars., variables; lim., limits.
\end{flushleft}
\endgroup
\end{table}

\subsubsection*{RSQ4. Design: How is algorithmic fairness addressed during the design phase of public health ML projects?}

In (S4.Q1) Stakeholder Diversity, 64\% reported moderate or stronger inclusion of diverse stakeholders, while one third reported minimal or none. Hospitals showed the highest levels of engagement, and universities the lowest. In (S4.Q2) Fairness Defined, 58\% indicated fairness was minimally or not defined, with fewer than 10\% reporting strong integration. Data scientists were more likely to mention explicit fairness definitions, while government respondents reported the weakest attention. In (S4.Q3) Data Diversity, 63\% reported moderate or stronger consideration of subgroup diversity, led by hospitals and public health researchers. In general, fairness considerations were more often implicit through stakeholder and data diversity than explicitly defined during design.

\begin{table}[h]
\centering
\caption{RSQ4: Design, response distribution from survey items S4.Q1–S4.Q3.}
\label{tab:rsq4_design}
\begingroup
\scriptsize
\setlength{\tabcolsep}{0pt}
\begin{tabular}{@{}p{2.9cm}p{1.2cm}p{2.9cm}p{1.2cm}p{2.9cm}p{1.2cm}@{}}
\toprule
\multicolumn{2}{c}{\parbox[c]{3.3cm}{\centering\textbf{(S4.Q1) Stakeholder Diversity}}} &
\multicolumn{2}{c}{\parbox[c]{3.3cm}{\centering\textbf{(S4.Q2) Fairness Defined}}} &
\multicolumn{2}{c}{\parbox[c]{3.3cm}{\centering\textbf{(S4.Q3) Data Diversity}}} \\
\cmidrule(lr){1-2}\cmidrule(lr){3-4}\cmidrule(l){5-6}
\textit{Answers} & \textit{n (\%)} &
\textit{Answers} & \textit{n (\%)} &
\textit{Answers} & \textit{n (\%)} \\
\midrule
Not at all & 3 (4) & Not at all & 10 (14) & Not at all & 7 (10) \\
Minimal & 23 (32) & Minimal & 32 (44) & Minimal & 20 (27) \\
Moderate & 33 (45) & Moderate & 25 (34) & Moderate & 29 (40) \\
Strong & 12 (16) & Strong & 5 (7) & Strong & 16 (22) \\
Fully & 2 (3) & Fully & 0 (0) & Fully & 1 (1) \\
\midrule
\multicolumn{2}{c}{{\textit{Total: }73 (100\%)}} &
\multicolumn{2}{c}{{\textit{Total: }73 (100\%)}} &
\multicolumn{2}{c}{{\textit{Total: }73 (100\%)}} \\
\bottomrule
\end{tabular}
\begin{flushleft}
\scriptsize
Percentages are rounded and may not total exactly 100\%.
\end{flushleft}
\endgroup

\end{table}

\subsubsection*{RSQ5. Application: How do researchers apply and justify trade-offs or interventions to operationalize fairness in practice?}

In (S4.Q5) Fairness Monitoring, 81\% reported no or minimal monitoring, and only 11\% indicated moderate or stronger activity. In (S5.Q1) Action Ranking, top priorities were ensuring recommendations were not worse for any income group (mean rank 2.12) and equal across groups (2.74), followed by equal recommendations for similar health needs (2.94) and prioritizing urgent needs (3.21); updating recommendations ranked lower (4.08). In (S5.Q2) Accuracy Trade-off, 84\% accepted some accuracy reduction for fairness, while 10\% rejected trade-offs and 6\% accepted a significant loss. In (S5.Q3) Bias Reporting, 72\% described or proposed improvements to bias findings, while 26\% implemented corrective actions. In general, respondents were open to modest accuracy–fairness trade-offs but rarely maintained systematic monitoring or corrective mechanisms.

\begin{table}[h]
\centering
\caption{RSQ5: Application, response distribution from survey items S4.Q5, S5.Q1-S5.Q3.}
\label{tab:rsq5_application}
\begingroup
\scriptsize
\setlength{\tabcolsep}{0pt}

\begin{tabular}{@{}p{2.2cm}p{1cm}p{2.8cm}p{1cm}p{2.25cm}p{1cm}p{2.5cm}p{1cm}@{}}
\toprule
\multicolumn{2}{c}{\parbox[c]{3.25cm}{\centering\textbf{(S4.Q5) Fairness Monitoring}}} &
\multicolumn{2}{c}{\parbox[c]{3.25cm}{\centering\textbf{(S5.Q1) Action Ranking}}} &
\multicolumn{2}{c}{\parbox[c]{3.25cm}{\centering\textbf{(S5.Q2) Accuracy Trade-off}}} &
\multicolumn{2}{c}{\parbox[c]{3.25cm}{\centering\textbf{(S5.Q3) Bias Reporting}}} \\
\cmidrule(lr){1-2}\cmidrule(lr){3-4}\cmidrule(lr){5-6}\cmidrule(l){7-8}
\textit{Answers} & \textit{n (\%)} &
\textit{Top priorities (mean rank)} & \textit{Score} &
\textit{Answers} & \textit{n (\%)} &
\textit{Answers} & \textit{n (\%)} \\
\midrule
Not at all & 41 (56) &
Equal across groups & 2.74 &
No reduction & 7 (10) &
No action & 0 (0) \\
Minimal & 18 (25) &
No group worse off & 2.12 &
Small & 45 (62) &
Acknowledge bias only & 0 (0) \\
Moderate & 6 (8) &
Similar needs & 2.94 &
Moderate & 16 (22) &
Suggest research & 28 (38) \\
Strong & 7 (10) &
Prioritize urgent needs & 3.21 &
Significant & 4 (6) &
Propose improvements & 25 (34) \\
Fully & 1 (1) &
Update over time & 4.08 &
Severe & 0 (0) &
Suggest corrections & 11 (15) \\
-- & -- &
Other (specify) & 6.00 &
-- & -- &
Immediate correction & 8 (11) \\
-- & -- & -- & -- & -- & -- & -- & -- \\
\midrule
\multicolumn{2}{c}{\textit{Total: }73 (100\%)} &
\multicolumn{2}{c}{\textit{Base: mean ranks per item (N = 64--68)}} &
\multicolumn{2}{c}{\textit{Total: }72 (100\%)} &
\multicolumn{2}{c}{\textit{Total: }72 (100\%)} \\
\bottomrule
\end{tabular}
\begin{flushleft}
\scriptsize
Percentages are rounded and may not total exactly 100\%. Mean ranks calculated as weighted averages of rank positions (1--6; lower = higher priority).
\end{flushleft}
\endgroup
\end{table}

\subsection{Theoretical Embedding}

\begin{figure}[ht]
    \centering
    \includegraphics[width=\linewidth]{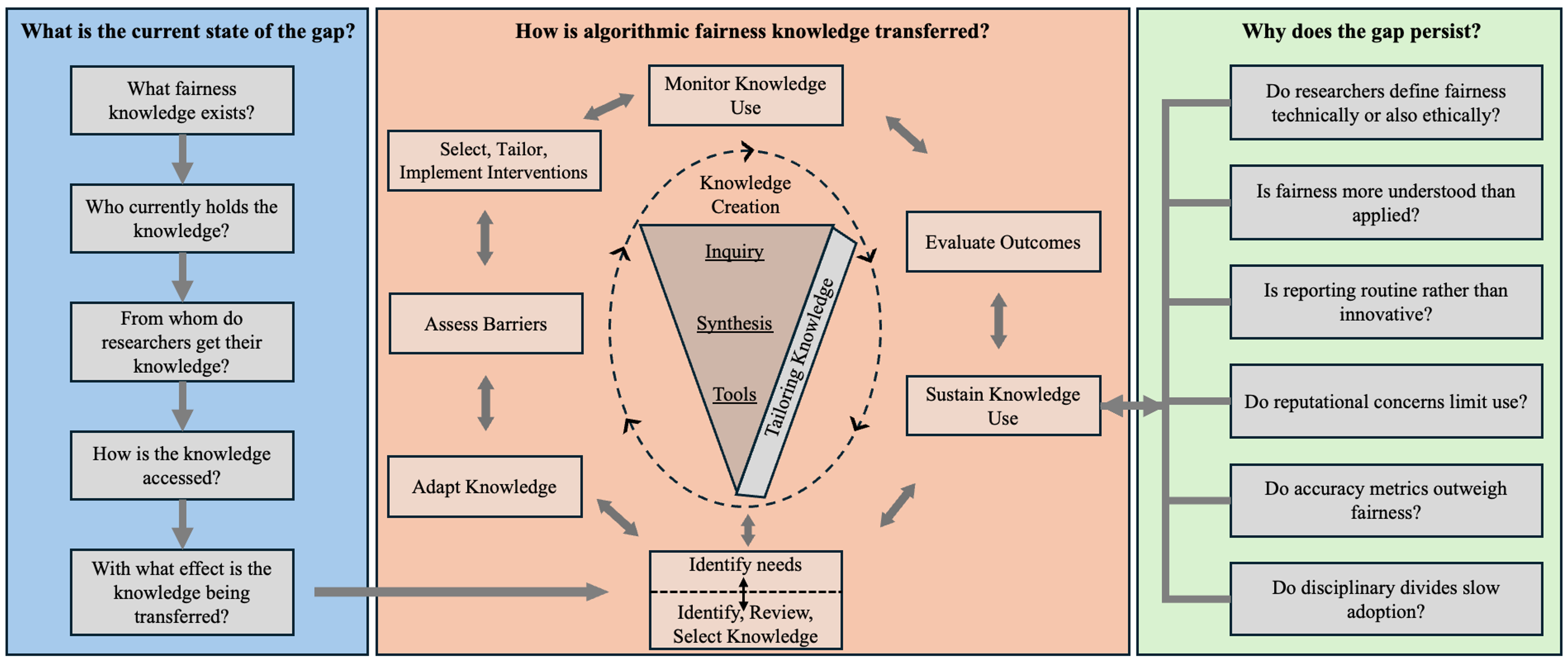}
    \caption{Fairness-to-Action (F2A) framework: analytical mapping of fairness translation across methodological (What), organizational (How), and systemic (Why) levels.}
    \label{fig:f2a_final}
\end{figure}

Survey findings were systematically mapped onto three theories on the research–practice gap: the Knowledge–Practice Gap (the What), the KTA Cycle (the How), and the Knowing–Doing Gap (the Why). Figure \ref{fig:f2a_final} presents the resulting F2A framework, composed of three panels mirroring these perspectives. The left panel lists diagnostic questions about the gap’s current state: what fairness knowledge exists, who holds it, where it is obtained, how it is accessed, and with what effect. The center panel shows KTA as two linked components—a knowledge-creation funnel (inquiry, synthesis, tools) and an action cycle (identify needs, select and adapt knowledge, assess barriers, implement, monitor, sustain). The right panel translates mechanisms from the Knowing–Doing Gap literature into diagnostic questions about why translation stalls, such as whether talk substitutes for action or accuracy metrics limit fairness. Horizontal connectors indicate flows from knowledge to action and feedback from sustainment to these explanatory mechanisms. Arrows are interpretive, depicting plausible feedbacks rather than causal claims. The framework offers an organizing scaffold consistent with observed patterns in ML-driven public health research and intended for use alongside implementation and governance guidance. Condensed mappings appear in Appendix \ref{appendix_condensed_mapping}; full matrices are available on request.


\paragraph{What is the current state of the gap?}
Regarding what fairness knowledge exists, respondents most often described fairness as equal treatment or protection of sensitive attributes, while smaller groups emphasized outcome-based definitions. Concerning who holds this knowledge, most reported limited understanding, with strong knowledge rare and particularly scarce in epidemiology and biostatistics, though higher levels appeared among data scientists and public health researchers. As to from whom researchers obtain it, dissemination relies mainly on external sources, while internal channels are weak or absent. With respect to how it is accessed, structured training opportunities are minimal, especially in universities and NGOs compared with hospitals. Finally, with what effect, reporting of algorithmic bias is typically minimal to moderate, with few comprehensive cases; reports often acknowledge bias or propose improvements rather than implement corrective actions. Acceptance of trade-offs between accuracy and fairness is widespread, though some respondents reject any accuracy reduction.

In summary, fairness knowledge exists but remains weakly institutionalized and not routine. It resides in individuals rather than organizations, depends on ad hoc external sources instead of structured training, and lacks feedback from reporting to remediation. Limitations are most pronounced where understanding is low, frameworks absent, and training missing.

\paragraph{How is algorithmic fairness knowledge transferred?}
In the knowledge-creation funnel, inquiry depends on external sources such as academic publications, conferences, and professional networks, while internal training channels remain weak. Synthesis is limited: most respondents reported minimal to moderate understanding, and framework awareness was common but often superficial, suggesting limited systematic integration of fairness knowledge. As for tools and products, frameworks or guidelines were rarely applied, and practices such as fairness audits or subgroup evaluation were inconsistent and limited in scope, with few examples of strong corrective action.

In the action cycle, problem recognition is partial: bias is occasionally observed, yet fairness reporting is minimal to moderate, indicating that issues are noticed but seldom documented. Selection and adaptation of knowledge are uneven: framework awareness varies by field, bias assessment is inconsistent, and attention to data diversity remains moderate, leaving subgroups under-considered. Barriers include technical and regulatory obstacles, limited ML expertise, and lack of training, with these gaps most evident in university and NGO contexts. Interventions are irregular—checking models for bias or evaluating performance across subgroups is common, while other actions such as involving stakeholders, publishing transparency reports, or integrating fairness into training are less frequent. Monitoring mechanisms are minimal, follow-up actions uncommon, and evaluation often framed as accuracy–fairness trade-offs without subsequent corrective steps.

Taken together, fairness knowledge transfer remains fragmented: inquiry relies on external channels, synthesis and tool use are superficial, and the action cycle stalls at implementation and monitoring.

\paragraph{Why does the gap persist?}
Regarding whether knowing what to do is sufficient, fairness was primarily defined in technical terms, with project-level definitions most often minimal or moderate, indicating limited integration of ethical or equity framings. On whether talk substitutes for action, understanding outpaced implementation: bias assessment and improvement practices were inconsistently applied across groups. Regarding whether memory substitutes for thinking, reporting was generally minimal or moderate, and uptake of new improvement practices remained uneven beyond routine activity. Concerning whether fear prevents acting on knowledge, bias was recognized, yet responses were cautious: although many accepted small or moderate accuracy reductions, follow-through into corrective action was limited. On whether measurement obstructs good judgment, bias mitigation was mostly none or minimal, while accuracy continued to dominate decision making. Finally, on whether internal competition fragments collective effort, differences across primary fields and institution types in ML use, training, and reporting indicated fragmentation rather than shared practice.

Overall, these results help explain why the research-practice gap persists: definitions remain narrow, understanding exceeds implementation, reporting is routine rather than innovative, accuracy concerns limit corrective action, fairness metrics lag behind accuracy measures, and disciplinary and institutional divides constrain collective progress.

\section{Discussion}

This study examined how researchers in ML-driven public health in the Netherlands perceive and operationalize algorithmic fairness. Extending earlier literature-based assessments of fairness reporting \citep{altamirano_machine_2025}, it provides an empirical view behind the scenes of how fairness is understood, where it is applied, and why translation often stalls within research practice.

Interpreted through the F2A framework, the findings reveal a multi-level challenge, with F2A serving as an evidence-informed diagnostic scaffold rather than a prescriptive or predictive model. At the methodological level (the What), fairness knowledge exists yet remains inconsistently defined, weakly institutionalized, and seldom formalized into design or reporting frameworks. At the organizational level (the How), knowledge flows primarily through external networks rather than internal structures, formal training is scarce, tools are applied unevenly and largely limited to subgroup checks or bias audits, and monitoring is minimal. At the systemic level (the Why), incentives continue to favor accuracy and throughput, reporting often substitutes for corrective action, and disciplinary as well as institutional divides fragment shared practice. Together, these conditions illustrate mechanisms that likely extend beyond the national setting, where fairness awareness often exceeds implementation.

The F2A framework helps explain where fairness translation falters: from defining and disseminating fairness knowledge to selecting and adapting it, from adaptation and barrier assessment to implementation, and from implementation and reporting to monitoring and sustained use. Each fragile transition reflects limited codification of fairness into design and reporting, weak organizational support, and incentives that undervalue iterative improvement. Rather than listing deficits, F2A highlights leverage points for strengthening translation: formalize definitions as design requirements, build internal knowledge pipelines and training, link reporting to corrective action, and align incentives with continuous monitoring and learning. Advancing fairness in practice therefore requires coordinated interventions across methodological, organizational, and systemic levels so that fairness becomes routine, evaluation continuous, and accountability shared. Algorithmic fairness in public health thus emerges not only as a methodological issue but also as an institutional and governance challenge. Sustained progress depends on infrastructures and accountability systems that anchor fairness as standard research practice rather than as an ad hoc or externally driven task.


\paragraph{Recommendations.}
We prioritize researcher- and group-level actions as most immediately feasible and high impact, while institutional, funding, and governance interventions represent longer-term, system-level levers. The three highest-impact near-term actions are to formalize fairness definitions as explicit design requirements, strengthen internal knowledge pipelines and training, and link bias reporting to concrete corrective action. At the individual and group levels, researchers can normalize fairness assessment alongside accuracy and robustness, cultivate shared accountability, reward fairness artifacts, and embed monitoring into routine workflows. Journals, conferences, and funding bodies can reinforce these practices by requiring fairness reporting, including subgroup evaluation, bias audits, and impact statements, and by supporting shared benchmarking datasets. Institutions and governments can enable sustained uptake through training, governance mechanisms, oversight, and standards that connect reporting to remediation. Finally, affected communities should be involved so that fairness definitions and metrics reflect lived experience rather than abstract proxies. Coordinated action across these levels targets the fragile transitions identified in F2A, helping to close the research-practice gap and ensure that ML-driven public health operates safely, transparently, and equitably, through feasible and coordinated action.

\section{Conclusion}
This study extends earlier assessments of algorithmic fairness in ML-driven public health by directly engaging researchers to examine how fairness is perceived and operationalized in practice. 
At a glance, the findings reveal three interlinked layers of translation failure: methodological (fairness knowledge is individually held and weakly codified); organizational (knowledge transfer relies on external rather than internal structures); and systemic (accuracy incentives and fragmented governance impede sustained fairness practice).
Framed through the F2A framework, these patterns highlight specific leverage points for strengthening translation: formalizing definitions as design requirements, developing internal knowledge pipelines and training, linking reporting to actionable remediation, and aligning incentives with ongoing monitoring and improvement. 
The contribution lies not in enumerating gaps but in providing a structured map of where interventions can anchor fairness as routine practice.

We acknowledge some limitations. The modest sample size limits statistical power for subgroup analyses, and self-reporting may introduce social desirability or recall bias. The national focus constrains generalizability, though the Netherlands offers a representative context for high-resource public health systems. In addition, simplifying complex definitions into single themes may obscure conceptual nuance. Finally, reported practices were not directly observed, so implementation could not be validated empirically.

Future work should address these limitations by validating fairness implementation through observed practice, tracing it across the ML lifecycle via case studies, and evaluating domain-specific metrics and mitigation methods. It should also test institutional levers (e.g., ethics review, funding criteria, training mandates) and examine how publication norms and governance shape reporting, develop lightweight tools for routine evaluation, and compare across countries to assess how structural conditions influence translation. These efforts can move the field from awareness to safer, sustained, accountable practice.

\section*{Acknowledgments} 
This research was conducted at the Civic AI Lab (SIAS group), Informatics Institute, University of Amsterdam, in collaboration with the Municipality of Amsterdam and the Dutch Ministry of the Interior and Kingdom Relations.

\bibliographystyle{unsrt}
\begingroup
\small
\bibliography{references}
\endgroup


\appendix
\section{Appendix}
\subsection{Full Survey Questionnaire}
\label{appendix_full_survey}
\footnotesize{
\subsubsection{Introduction}
Welcome, and thank you for participating in this survey on Algorithmic Fairness in Dutch Public Health Research. 
This survey is part of a larger research study conducted by the Civic AI Lab at the University of Amsterdam. 
The study examines the use of Machine Learning in Dutch Public Health Research, focusing on algorithmic fairness. 
Following a literature review, this survey engages researchers directly to provide deeper insights into how algorithmic bias is conceptualized, assessed, and mitigated in practice. 
While the literature offers valuable insights, firsthand perspectives are essential to capture the rationale behind research decisions. 
The survey consists of five sections, and it will take up to 15 minutes to complete.  

To ensure the reliability and robustness of responses, we kindly request your email address. To thank you for your participation, we are offering a raffle of five Hema €10 gift cards. Email addresses will only be used to notify winners of the raffle and will not be shared or used for any other purposes.  

\textbf{Key Definitions:}  
\begin{itemize}
    \item Algorithmic Decision Making: The use of Machine Learning to make automated decisions based on data, such as predicting health outcomes or allocating resources in Public Health.  
    \item Algorithmic Bias: Systematic and unfair discrimination that arises when Machine Learning models produce prejudiced outcomes.  
    \item Algorithmic Fairness: The practice of ensuring that decisions made by machines are fair and do not favor certain groups over others.  
\end{itemize}

\subsubsection{Consent}
\textit{Informed Consent:}  
By proceeding, you confirm that:  
\begin{itemize}
    \item You understand the purpose of this study, and your participation is voluntary.  
    \item You agree that your anonymized responses may be used for research purposes.  
    \item You acknowledge that you can withdraw at any time without consequences.  
\end{itemize}

\subsubsection{Section 0: Demographics}
\textit{This section gathers information about your background to help us understand the diversity of perspectives in the survey. All questions in this section are required.}

\begin{description}
    \item[S0.Q1] What is your primary field of expertise?  
    \begin{itemize}
        \item Public Health Research
        \item Data Science / Machine Learning
        \item Epidemiology / Biostatistics
        \item Health Policy \& Management
        \item Other (please specify)
    \end{itemize}

    \item[S0.Q2] What is your current role?  
    \begin{itemize}
        \item Researcher / Academic
        \item Data Scientist / Machine Learning Engineer
        \item Healthcare / Public Health Professional
        \item Policy Maker / Government Official
        \item Other (please specify)
    \end{itemize}

    \item[S0.Q3] How many years of experience do you have in Public Health research?  
    \begin{itemize}
        \item Less than 1 year
        \item 1--3 years
        \item 4--7 years
        \item 8--10 years
        \item More than 10 years
    \end{itemize}

    \item[S0.Q4] Have you worked with Machine Learning methods in your research, either directly or through collaboration?  
    \textit{Includes using Machine Learning for tasks like disease prediction, patient classification, or co-authoring Machine Learning-based studies.}
    \begin{itemize}
        \item Yes, extensively
        \item Yes, occasionally
        \item No, but I plan to
        \item No, and I do not plan to
        \item Not sure
    \end{itemize}

    \item[S0.Q5] In what type of institution do you primarily conduct research?  
    \begin{itemize}
        \item University / Academic Institution
        \item Government or Public Health Organization
        \item Hospital or Healthcare Institution
        \item Private Research Institution
        \item Non-Governmental Organization (NGO)
        \item Other (please specify)
    \end{itemize}
\end{description}

\subsubsection{Section 1: Extent of Usage of Algorithmic Decision-Making}
\textit{Let’s begin by exploring your experience with Machine Learning in Public Health research.}

\begin{description}
    \item[S1.Q1] What is your understanding of algorithmic decision-making (i.e., using Machine Learning to make data-based decisions)? [Open text response]  

    \item[S1.Q2] To what extent is Machine Learning used in your research, either directly or through collaboration?  
    \textit{Consider any instance where you have used computational models to analyze health data, classify cases, or predict outcomes.}
    \begin{itemize}
        \item Rarely
        \item Occasionally
        \item Sometimes
        \item Frequently
        \item Extensively
    \end{itemize}

    \item[S1.Q3] What key factors influence your decision to use Machine Learning in your research?  
    \begin{itemize}
        \item Costs of model development
        \item Time to deployment
        \item Scalability
        \item Applicability across different groups
        \item Ethical considerations
        \item Data quality and availability
        \item Regulatory / compliance requirements
        \item Other (please specify)
    \end{itemize}

    \item[S1.Q4] What challenges have you faced when using Machine Learning in your research?  
    \begin{itemize}
        \item Limited access to high-quality data
        \item Limited computing power
        \item Legal / regulatory barriers
        \item Difficulties with interoperability or data sharing
        \item Lack of Machine Learning expertise within the research team
        \item Concerns about model transparency and explainability
        \item Other (please specify)
    \end{itemize}
\end{description}

\subsubsection{Section 2: Algorithmic Fairness Awareness}
\textit{Now, we’ll look at your knowledge and practices related to algorithmic fairness in Machine Learning.}

\begin{description}
    \item[S2.Q1] How would you define algorithmic fairness? [Open text response]  

    \item[S2.Q2] To what extent is algorithmic fairness considered and understood in your (research) group or organization?  
    \textit{Consider both formal policies and informal discussions on fairness in Machine Learning applications.}
    \begin{itemize}
        \item Not at all
        \item Minimally
        \item Moderately
        \item Strongly
        \item Fully
    \end{itemize}

    \item[S2.Q3] Are you aware of any existing guidelines, policies, or frameworks within your group, organization, or field to address algorithmic fairness?  
    \begin{itemize}
        \item Yes, within our group/organization
        \item Yes, but only in our field
        \item Yes, aware but no details
        \item No, none exist
        \item Not sure
        \item Other (please specify)
    \end{itemize}

    \item[S2.Q4] How do you stay updated on the latest developments in algorithmic fairness?  
    \begin{itemize}
        \item Academic networks and research initiatives
        \item Peer-reviewed articles and repositories
        \item Conferences and workshops
        \item Social media and online networks
        \item Internal training on algorithmic fairness
        \item I do not stay updated
        \item Other (please specify)
    \end{itemize}

    \item[S2.Q5] Are there any ongoing formal training or educational programs on algorithmic fairness in your group or organization?  
    \begin{itemize}
        \item Yes, regular formal training programs
        \item Yes, periodic workshops or seminars
        \item Researchers pursue external training independently
        \item No, but we plan to start training in the near future
        \item No, we do not have any training programs
        \item Other (please specify)
    \end{itemize}
\end{description}

\subsubsection{Section 3: Algorithmic Fairness in Practice}
\textit{Next, let’s dive into how biases are assessed and addressed in Machine Learning applications.}

\begin{description}
    \item[S3.Q1] How does your research group assess biases that may arise from the use of Machine Learning?  
    \begin{itemize}
        \item Sensitivity analysis of underrepresented groups
        \item Evaluating the impact of demographic and social variables
        \item Focus groups with healthcare workers or stakeholders
        \item Informal discussions
        \item Advanced statistical methods (fairness metrics, bias detection)
        \item Independent reviews by external experts
        \item No measures in place
        \item Other (please specify)
    \end{itemize}

    \item[S3.Q2] In your experience, have you observed any of the following types of algorithmic bias in Public Health research?  
    \begin{itemize}
        \item Bias in prediction accuracy across demographic groups
        \item Disparities in health recommendations based on socioeconomic status
        \item Data quality issues leading to biased outcomes
        \item Bias in model validation due to sample diversity
        \item Algorithmic bias in resource allocation
        \item Discrimination based on underrepresented groups in training data
        \item I have not observed algorithmic bias
        \item Other (please specify)
    \end{itemize}

    \item[S3.Q3] To what extent does your research group report on algorithmic bias in scientific studies?  
    \begin{itemize}
        \item Not at all
        \item Minimal
        \item Moderate
        \item Strong
        \item Fully
    \end{itemize}

    \item[S3.Q4] What steps have you implemented to promote fairness, transparency, and accountability in Machine Learning studies?  
    \begin{itemize}
        \item Open sharing of code
        \item Establishing advisory boards
        \item Involving stakeholders in the decision-making process
        \item Public availability of algorithms
        \item Customized explanation approaches
        \item Regular fairness audits and assessments
        \item Internal or external transparency reports
        \item None
        \item Other (please specify)
    \end{itemize}
\end{description}

\subsubsection{Section 4: Current Practices in Machine Learning}
\textit{Please answer the following questions based on a specific Machine Learning application or project you have worked on. Consider the design, development, and implementation of that project when answering.}

\begin{description}
    \item[S4.Q1] How diverse were the stakeholders involved in the design and development of the algorithm?  
    \begin{itemize}
        \item Not at all
        \item Minimal
        \item Moderate
        \item Strong
        \item Fully
    \end{itemize}

    \item[S4.Q2] How well-defined was the concept of algorithmic fairness for your application, including awareness of potential biases in data and models?  
    \begin{itemize}
        \item Not at all
        \item Minimal
        \item Moderate
        \item Strong
        \item Fully
    \end{itemize}

    \item[S4.Q3] To what extent did you ensure and assess the diversity of your training and evaluation data, across different subgroups and locations?  
    \begin{itemize}
        \item Not at all
        \item Minimal
        \item Moderate
        \item Strong
        \item Fully
    \end{itemize}

    \item[S4.Q4] How comprehensively did you evaluate and mitigate bias in your algorithm, including the use of fairness metrics and corrective measures?  
    \textit{Fairness metrics are tools to assess if models treat groups equally.}
    \begin{itemize}
        \item Not at all
        \item Minimal
        \item Moderate
        \item Strong
        \item Fully
    \end{itemize}

    \item[S4.Q5] To what extent do you have mechanisms for ongoing fairness monitoring and stakeholder training on bias?  
    \begin{itemize}
        \item Not at all
        \item Minimal
        \item Moderate
        \item Strong
        \item Fully
    \end{itemize}
\end{description}

\subsubsection{Section 5: Fairness in Action}
\textit{In this section, you will be presented with a practical scenario about a health app, where you'll consider fairness decisions and trade-offs based on your experience.}

\textbf{Scenario: Health App Recommendations (Income Bias).} Imagine designing a Public Health app with personalized lifestyle recommendations. After reviewing its performance, you find that lower-income neighborhoods receive fewer recommendations than wealthier ones (i.e., unequal access).

\begin{description}
    \item[S5.Q1] Please rank the following bias-mitigating actions from most to least important (1 = top priority):  
    \begin{enumerate}
        \item Ensure all income groups receive equal recommendations (equality)
        \item Ensure recommendations are not worse for any income group (equity)
        \item Ensure people with similar health needs receive similar recommendations
        \item Prioritize the most urgent health needs, even if recommendations are not equally distributed
        \item Regularly update recommendations to improve fairness over time
        \item Other (please specify)
    \end{enumerate}

    \item[S5.Q2] Would you be willing to reduce accuracy in the app’s recommendations to ensure equal access to all income groups?  
    \begin{itemize}
        \item No reduction in accuracy
        \item Small reduction
        \item Moderate reduction
        \item Significant reduction
        \item Severe reduction
    \end{itemize}

    \item[S5.Q3] How would you report the discovered bias against lower-income users in your scientific paper?  
    \begin{itemize}
        \item Decide not to address the bias
        \item Acknowledge the bias without further action
        \item Mention the bias and suggest further research
        \item Explain the bias and propose future improvements
        \item Provide a detailed analysis and suggest corrective actions
        \item Provide a detailed analysis and implement corrective actions immediately
        \item Other (please specify)
    \end{itemize}
\end{description}
}

\subsection{Condensed Theoretical Mapping}
\label{appendix_condensed_mapping}
\scriptsize
\setlength{\tabcolsep}{2pt}

\subsubsection{Knowledge–Practice Gap (What)}
\begin{tabular}{p{2.5cm} p{2.7cm} p{2.7cm} p{3.8cm} p{1.3cm}}
\toprule
\textbf{Original question} & \textbf{Definition (from source)} & \textbf{Adapted to fairness in public health ML} & \textbf{Our adapted question} & \textbf{Mapped to survey items} \\
\midrule
What should be transferred? & Moving research knowledge into routine practice & Fairness knowledge exists but is diffuse or implicit & How do researchers define and use fairness concepts in practice? & S2.Q1, S2.Q2 \\
\addlinespace[2pt]
Whose knowledge should be transferred? & Source may be individuals, teams, or organizations & Identify who holds and conveys fairness knowledge & Who conveys fairness knowledge within projects and institutions? & S2.Q4 \\
\addlinespace[2pt]
To whom should knowledge be transferred? & Target audiences vary by role and setting & Determine recipients of fairness knowledge & Which stakeholders receive fairness knowledge and guidance? & S4.Q1 \\
\addlinespace[2pt]
How should knowledge be transferred? & Mechanisms include guidelines, policies, systems & Organizational and informal mechanisms for fairness & What mechanisms support the transfer of fairness knowledge? & S3.Q1, S3.Q3 \\
\addlinespace[2pt]
What is the context for transfer? & Local settings shape feasibility and uptake & Public health settings and constraints influence fairness efforts & What contextual factors enable or hinder fairness transfer? & S1.Q4, S4.Q3 \\
\addlinespace[2pt]
What effects should be measured? & Outcomes should reflect change in practice & Effects include bias detection, mitigation, and reporting & How is the effect of fairness efforts assessed and reported? & S3.Q2, S4.Q4 \\
\bottomrule
\end{tabular}

\vspace{0.6em}

\subsubsection{Knowledge to Action Cycle (How)}
\begin{tabular}{p{2.5cm} p{2.7cm} p{2.7cm} p{3.8cm} p{1.3cm}}
\toprule
\textbf{Original component} & \textbf{Definition (from source)} & \textbf{Adapted to fairness in public health ML} & \textbf{Our adapted question} & \textbf{Mapped to survey items} \\
\midrule
Identify the problem & Define a gap that needs addressing & Define fairness gaps relevant to the project & What fairness problems are recognized at project start? & S1.Q3, S4.Q2 \\
\addlinespace[2pt]
Select knowledge & Find and appraise applicable knowledge & Locate fairness frameworks, metrics, and guidance & From which sources is fairness knowledge drawn? & S2.Q4, S2.Q3 \\
\addlinespace[2pt]
Adapt to local context & Tailor knowledge to setting and needs & Adapt fairness guidance to public health data and stakeholders & How is fairness adapted to the project and setting? & S4.Q2, S4.Q3 \\
\addlinespace[2pt]
Assess barriers & Identify barriers and facilitators & Practical, organizational, and data-related barriers to fairness & What barriers hinder fairness activities? & S1.Q4, S3.Q1 \\
\addlinespace[2pt]
Select and implement interventions & Choose and apply actions & Choose fairness evaluations, mitigations, and reporting & Which fairness actions are implemented and how? & S4.Q4, S3.Q4 \\
\addlinespace[2pt]
Monitor use & Track application over time & Ongoing monitoring and training on fairness & How is fairness use monitored and sustained? & S4.Q5 \\
\addlinespace[2pt]
Evaluate outcomes & Measure impact on outcomes & Evaluate bias, equity, and transparency outcomes & How are outcomes of fairness actions evaluated? & S3.Q2, S3.Q3 \\
\bottomrule
\end{tabular}

\vspace{0.6em}

\subsubsection{Knowing–Doing Gap (Why)}
\begin{tabular}{p{2.5cm} p{2.7cm} p{2.7cm} p{3.8cm} p{1.3cm}}
\toprule
\textbf{Original mechanism} & \textbf{Definition (from source)} & \textbf{Adapted to fairness in public health ML} & \textbf{Our adapted question} & \textbf{Mapped to survey items} \\
\midrule
Knowing “what” is not enough & Knowledge does not ensure action & Fairness is defined but not operationalized & Why does fairness knowledge fail to become practice? & S2.Q1, S3.Q1 \\
\addlinespace[2pt]
Talk substitutes for action & Discussion replaces implementation & Fairness is acknowledged but actions are limited & Does discussion of fairness replace mitigation or monitoring? & S3.Q3, S4.Q4 \\
\addlinespace[2pt]
Memory substitutes for thinking & Habit outweighs evidence-based change & Legacy practices persist over fairness updates & Do teams rely on precedent instead of updating with fairness guidance? & S4.Q2, S4.Q3 \\
\addlinespace[2pt]
Fear prevents acting on knowledge & Perceived risks block change & Concerns about performance or compliance stall fairness actions & Do perceived risks prevent implementing fairness steps? & S1.Q3, S1.Q4 \\
\addlinespace[2pt]
Measurement obstructs judgment & Metrics overshadow good decisions & Narrow metrics crowd out broader fairness goals & Do chosen metrics hinder broader fairness judgments? & S4.Q4, S5.Q2 \\
\bottomrule
\end{tabular}
\normalsize


\subsection{Codebooks for Algorithmic Decision-Making and Algorithmic Fairness}
\scriptsize
\setlength{\tabcolsep}{3pt}

\subsubsection*{Codebook: Algorithmic Decision-Making (ADM)}
\begin{tabular}{p{0.5cm} p{1.2cm} p{2cm} p{5.2cm} p{4cm}}
\toprule
\textbf{ID} & \textbf{Short Label} & \textbf{Theme} & \textbf{Description} & \textbf{Keywords} \\
\midrule
D & Data & Data Analysis & ADM framed as analyzing or processing data to generate outputs, insights, or conclusions, emphasizing extraction, discovery, or mining. & analyze, analysis, processing, extract, discovery, mining, output, insight, evaluation, conclusion, suggestion \\
\addlinespace[2pt]
M & Model & Model Focus & ADM defined in terms of models or algorithms themselves, highlighting their central role in predictions or decisions. & model, models, algorithm, methods, logistic regression, decision tree \\
\addlinespace[2pt]
R & Rules & Rules-Based & ADM described as following predefined rules, formulas, or statistical inference, with decisions derived from fixed logic. & rule, rules, formula, predefined, statistical, inference \\
\addlinespace[2pt]
P & Pattern & Pattern Recognition & ADM understood as recognizing patterns, trends, or regularities in data, including classification or categorization. & pattern, recognition, classify, classification, clustering, trends \\
\addlinespace[2pt]
S & Support & Decision Support & ADM portrayed as assisting human decision-making through guidance, recommendations, or scores, while leaving the final choice to humans. & assist, support, guide, augment, recommendation, score, outcome \\
\addlinespace[2pt]
G & Governance & Governance \& Fairness & ADM framed in terms of values, risks, or ethical dimensions, including fairness, bias, transparency, accountability, and limitations. & fairness, bias, transparency, black box, limitation, accountability \\
\bottomrule
\label{tab:codebook_adm}
\end{tabular}

\vspace{1em}

\subsubsection*{Table B: Algorithmic Fairness Codebook}
\begin{tabular}{p{0.5cm} p{1.2cm} p{2cm} p{5.2cm} p{4cm}}
\toprule
\textbf{ID} & \textbf{Short Label} & \textbf{Theme} & \textbf{Description} & \textbf{Keywords} \\
\midrule
P & Protected & Protected Attributes & Fairness means avoiding bias or discrimination specifically related to sensitive or protected attributes such as race, gender, age, or socioeconomic status. & race, gender, age, ethnicity, minority, socioeconomic, subgroup, protected \\
\addlinespace[2pt]
O & Outcome & Outcome Fairness & Fairness is described as ensuring predictions, classifications, or outcomes are accurate, consistent, or balanced across groups, without explicit reference to attributes. & outcome, result, accuracy, prediction, classification, balance, consistency \\
\addlinespace[2pt]
S & Standards & Standards \& Processes & Fairness is defined through technical, procedural, or governance measures, including unbiased or representative data, debiasing, fairness constraints, monitoring, and compliance. & dataset, representative, debias, constraints, governance, ethics, transparency, monitoring, audit \\
\addlinespace[2pt]
C & Context & Context \& Accountability & Fairness is linked to broader social or institutional context, accountability, or responsibility, including preventing reinforcement of social inequalities. & accountability, responsibility, social, inequality, inequities, context, institutional \\
\addlinespace[2pt]
E & Equality & Equal Treatment & Fairness is defined in generic terms of treating everyone the same, avoiding bias or discrimination, and ensuring equal treatment without group-specific references. & fair, equal, equality, nondiscrimination, impartial, justice \\
\addlinespace[2pt]
D & Disparities & Equity \& Disparities & Fairness framed as addressing inequities or disparities, ensuring non-inferior care, avoiding harm to disadvantaged groups, and promoting equity in outcomes. & healthcare, patient, disparities, underserved, equity, inequities, harm, safety \\
\bottomrule
\label{tab:codebook_af}
\end{tabular}

\vspace{1em}

\normalsize
\paragraph{Coding Rules}
\begin{itemize}
    \item Assign each response to exactly one theme.  
    \item Match to theme keywords where possible; if none, use conceptual match (main idea).  
    \item Apply human judgment when wording is indirect.  
    \item Themes are mutually exclusive; select the strongest idea.  
    \item All valid responses are coded; “nan” entries remain uncoded.  
\end{itemize}


\subsection{Ethical Considerations}\label{appendix:ethics}

Separate consent procedures were implemented for each phase. Interview participants received an information sheet and provided signed consent before participation. Survey participants reviewed an online consent statement and had to provide consent before proceeding. Participation was voluntary and withdrawal was possible at any time without consequence.
To incentivize participation, survey respondents could enter a raffle for five \texteuro10 gift cards. Email addresses were collected solely for validation and prize notification, stored in a file separate from survey data, and deleted after notification. The raffle file contained no survey variables and could not be used to re-identify responses.
Data protection complied with GDPR and University of Amsterdam guidance. Interview data were stored on SURFdrive (a secure, GDPR-compliant institutional platform). Survey data were stored in a password-protected institutional environment with access limited to the research team. Before analysis, the dataset was anonymized: direct identifiers (names, email addresses) and high-risk quasi-identifiers (institutional domains, IP addresses, precise timestamps) were removed. No linkage keys or auxiliary files were retained, preventing any re-linkage. Responses to “Other (please specify)” were checked to ensure they contained no identifying information. The study involved no sensitive personal data, no medical or invasive procedures, and no vulnerable populations. All participants provided free, informed, and unambiguous consent. 

\end{document}